\documentclass[aps,prev,twocolumn,preprintnumbers,floatfix,nofootinbib]{revtex4-1}
\pdfoutput=1

\newcommand{\be}{\begin{equation}}
\newcommand{\ee}{\end{equation}}
\newcommand{\bea}{\begin{eqnarray}}
\newcommand{\eea}{\end{eqnarray}}

\usepackage{aas_macros}

\usepackage{mathrsfs}
\usepackage{amsmath, amsthm, amssymb}
\usepackage{color}
\usepackage{epstopdf}
\usepackage[pdftex]{graphicx}
\usepackage{amssymb}
\usepackage{verbatim}
\usepackage{hyperref}
\usepackage{enumerate}
\usepackage{float}
\usepackage[caption = false]{subfig}
\usepackage[normalem]{ulem}  

\usepackage[T1]{fontenc}

\usepackage[normalem]{ulem}
\usepackage{relsize}
\usepackage{mathptmx}
\usepackage{color}
\usepackage{xcolor}

\begin{document}

%\title{On Formally Undecidable Propositions in Gauge Theory}
\title{Measuring the Quantum State of Dark Matter}
%\title{The Taming of the Axion}

\author{David J. E. Marsh$^{a}$}
\email{david.j.marsh@kcl.ac.uk}

\vspace{1cm}
\affiliation{${}^a$ Theoretical Particle Physics and Cosmology, King's College London, Strand, London, WC2R 2LS, United Kingdom}

\begin{abstract}

I demonstrate a simple example of how the time series obtained from searches for ultralight bosonic dark matter (DM), such as the axion, can be used to determine whether it is in a coherent or incoherent quantum state. The example is essentially trivial, but I hope that explicitly addressing it provokes experimental exploration. In the standard coherent state, $\mathcal{O}(1)$ oscillations in the number density occur over the coherence time, $\tau_c=h/m v^2$, where $m$ is the particle mass and $v$ is the galactic virial velocity, leading to a reduction in the constraining power of experiments operating on timescales $T<\tau_c$, due to the unknown global phase. On the other hand if the DM is incoherent then no such strong number oscillations occur, since the ensemble average over particles in different streams gives an effective phase average. If an experiment detects a signal then the coherent or incoherent nature of DM can be determined by time series analysis over the coherence time. This finding is observationally relevant for DM masses, $10^{-17}\text{ eV}\lesssim m\lesssim 10^{-11}\text{ eV}$ (corresponding to coherence times between a year and 100 seconds), and can be explored by experiments including CASPEr, DMRadio, and AION. Coherence may also be measurable at higher masses in the microwave regime, but I have not explored it.

\end{abstract}

\maketitle
%%%%%%%%%%%%%%%%%%%%%%%%%%%%%%%%%%%%%%%%%%%%%%%%%%%%%%%%%%%%%%%%
%%%%%%%%%%%%%%%%%%%%%%%%%%%%%%%%%%%%%%%%%%%%%%%%%%%%%%%%%%%%%%%%

\section{Introduction} 

The quantum nature of dark matter (DM) has important observational consequences. If the particle mass, $m$, is less than a few hundred eV, then the Pauli exclusion principle applied to galactic halos implies that the DM must necessarily be bosonic~\cite{1979PhRvL..42..407T,Alvey:2020xsk}. If the mass is less than around 1 eV, then the occupation number of the local dark matter density, $\rho_{\rm loc}\approx 0.4\text{ GeV cm}^{-3}$~\cite{ParticleDataGroup:2020ssz}, within a de Broglie wavelength is large, allowing for description in terms of a classical field $\phi$, according to well-known arguments~\cite{sakurai}. Such ultralight bosonic DM (UBDM) is the subject of this \emph{perspective}.

A prototypical model of UBDM is provided by the axion~\cite{pecceiquinn1977,weinberg1978,wilczek1978}. Axion DM can be created in the early Universe by a process known as vacuum realignment~\cite{1983PhLB..120..127P,1983PhLB..120..133A,1983PhLB..120..137D} from an initial supercooled, squeezed~\cite{Kuss:2021gig}, and highly homogeneous state that arises during cosmic inflation. This mechanism also functions for any real scalar field DM with a quadratic potential minimum~\cite{1983PhRvD..28.1243T}. The subsequent evolution of the DM field is then typically found~\cite{Marsh:2015xka} by solving the classical mean field theory (MFT) equations, in both the linear, non-linear, and even fully relativistic regimes (see e.g. Refs.~\cite{Hlozek:2014lca,Schive:2014dra,2017JCAP...03..055H} respectively, and Ref.~\cite{Gorghetto:2022sue} for the same phenomena with ultralight vector DM). %This description is very briefly reviewed in the Supplementary Material. 

The mean field description includes a variety of wavelike effects, which have important observational consequences and distinguishes UBDM from other DM models (such as cold DM, warm DM, or any models based on heavy fermions such as weakly interacting massive particles). Such wave phenomena arise due to the gradient energy in the field, and interference. At the linear level, gradient energy leads to an effective pressure and Jeans scale~\cite{khlopov_scalar}, reducing the gravitational clustering of UBDM on small scales~\cite{2010PhRvD..82j3528M}, which is a cosmic manisfestation of the de Broglie wavelength. In the non-linear, multistreaming regime after shell crossing, interference effects begin to manifest, first in cosmic filaments, and later inside DM halos (e.g. Refs.~\cite{Schive:2014dra,Veltmaat:2018dfz,Lague:2020htq,Gough:2022pof} ). 

\begin{figure}[!t]
\centering
\includegraphics[width=0.9\linewidth]{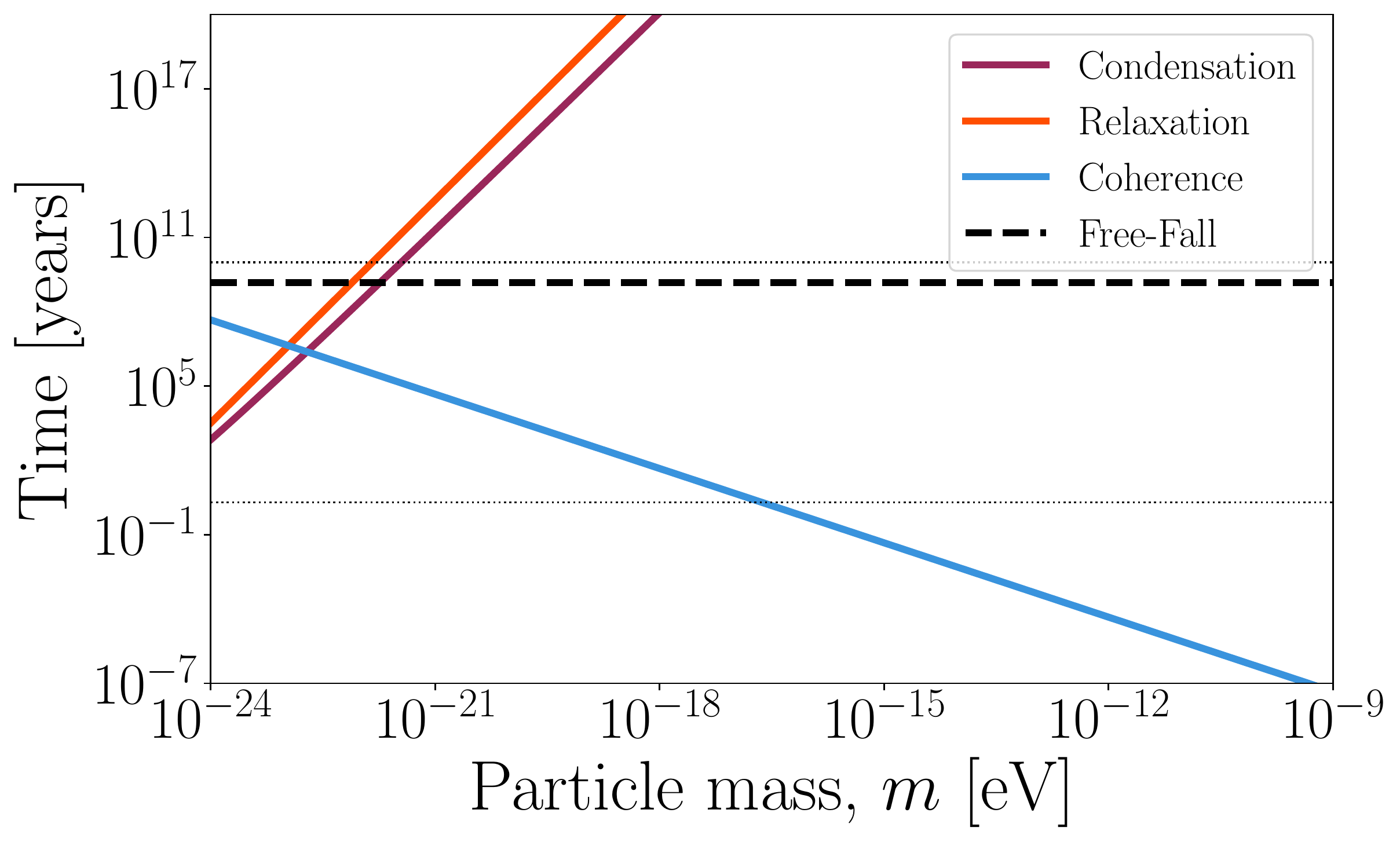}
\caption{Typical time scales for coherent wave phenomena in the Milky Way. Horizontal dotted lines indicate 1 year and the age of the Universe, respectively. Coherence can be observed in an experiment if the timescale is shorter than around one year, implying $m\gtrsim 10^{-17}\text{ eV}$. In this regime the condensation and relaxation times are much longer than the age of the Universe.}  
\label{fig:timescales}
\end{figure}

Inside DM halos, UBDM virialises with a characteristic velocity $v_0$, which is of order 200 km s$^{-1}$ in the Milky Way. The classical field model then leads to the presence of coherence patches of size $L_c\approx 1/mv_0$ (I use units $\hbar=c=1$ in most of the following), which are stable on the coherence time $\tau_c\approx1/mv_0^2$. The dynamics of these coherence patches has the effect of heating and cooling stars by two body relaxation~\cite{BinneyTremaine2008} on time scale $\tau_{\rm rel.}$, which has been used to place constraints on the allowed UBDM particle mass (e.g. Refs.~\cite{Hui:2016ltb,Marsh:2018zyw}). Finally, wavelike effects lead to the formation of self-bound non-linear objects known as boson stars, which form either by violent relaxation during virialisation~\cite{Schive:2014dra,Veltmaat:2018dfz} on the gravitational free fall time, $\tau_{\rm ff}$, or due to gravitational Bose-Einstein condensation in the kinetic regime~\cite{Levkov:2018kau}, with condensation time $\tau_{\rm cond.}$ (note that this phenomenon is \emph{classical} wave condensation~\cite{2012NatPh...8..471S}). These timescales are shown in Fig.~\ref{fig:timescales}. 

In order for coherence to be measurable, $\tau_c$ should be shorter than the experimental timescale. Assuming a reasonable campaign of $\mathcal{O}(1)\text{ year}$ implies that coherence can be measured only for $m\gtrsim 10^{-17}\text{ eV}$. We see that in this case, the relaxation and condensation times in the Milky Way are much longer than the age of the Universe, and thus are not dynamically relevant.~\footnote{The coincidence of these three timescales around $m\approx 10^{-23}\text{ eV}$, combined with coherence and coondensation times shorter than the age of the Universe, can be thought of as defining the ``Fuzzy Dark Matter'' regime of galactic phenomenology, reviewed in e.g. Refs.~\cite{Niemeyer:2019aqm,Hui:2021tkt}.}

%%%%%%%%%%%%%
%%%%%%%%%%%%%%%%

There are two quantum timescales to consider: the \emph{quantum break time}, and the \emph{decoherence timescale}. Ref.~\cite{Allali:2020ttz} computed the gravitational decoherence time, which becomes shorter than the age of the Universe for $m\lesssim 7\times 10^{-7}\text{ eV}$. The decoherence time scale is shorter than all timescales shown in Fig.~\ref{fig:timescales} over the particle mass range plotted. 

The quantum break time is considered by Ref.~\cite{Eberhardt:2022rcp}, who have  argued that beyond MFT effects can become important on just a few times the free-fall time, $\tau_{\rm ff}$, despite the huge occupation numbers of UBDM, leading to shorter quantum break times (for other beyond MFT studies, see Ref.~\cite{Lentz:2019xcr}). It is argued that a short quantum break time leads to an \emph{incoherent} description of the local DM density due to the mixing of different streams in which the initially coherent phase diverges, akin to classical chaos. Particles with fixed energy acquire different quantum phases, and coherent wavelike effects are hypothesised to be suppressed. Given the importance of coherence effects in constraining the properties of UBDM using astrophysics and cosmology, it is worth asking whether this can be tested in the laboratory. It is this question that I explore in the following.

\section{A simple coherent field model} 
%%%%%%%%%%%%%
\begin{figure*}[!t]
\centering
\includegraphics[width=1.5\columnwidth]{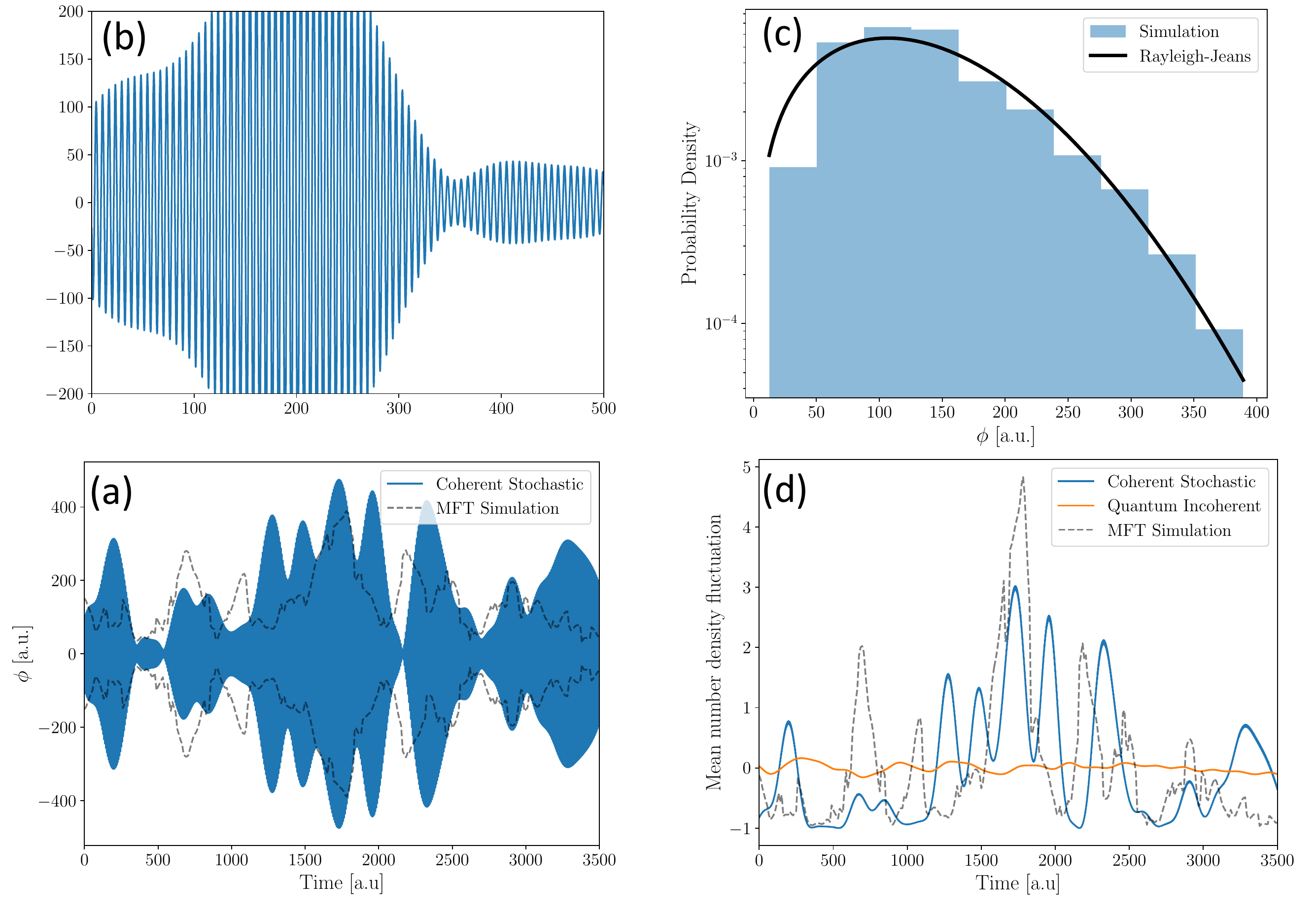}
\caption{(a) Stochastic model for the coherent UBDM field and comparison to MFT simulation of a cosmological UBDM halo from Ref.~\cite{Veltmaat:2018dfz} (arbitrarily normalised). The MFT simulation is in the WKB limit, only giving the envelope function, which varies on the coherence time, $1/mv_0^2$, with no rapid oscillations on timescale $1/m$. (b) Insert showing rapid oscillations in the stochastic model. (c) Rayleigh-Jeans distribution of the field amplitude, and histogram from MFT simulations. (d) Relative number density fluctuations in the coherent stochastic model, the quantum incoherent model with 200 streams, and MFT simulations. Note that the stochastic model is (roughly) normalised to the MFT amplitude and coherence time, but by nature does not fit the exact time evolution. }  
\label{fig2}
\end{figure*}
%%%%%%%%%%%%%%%%

The local DM field can be modelled stochastically as~\cite{Krauss:1985ub,Derevianko:2016vpm,Foster:2017hbq}:
\be
\phi(t) =\frac{\sqrt{2\rho_{\rm loc}}}{m} \sum_{i=0}^{i_{\rm max}} \alpha_i \cos \left[ \left(1+\frac{v_i^2}{2}\right)mt+\delta_i\right]\, ,
\label{eqn:coherent}
\ee
where $\delta_i\in \mathcal{U}[-\pi,\pi]$ is a random phase and $\alpha_i$, and $v_i$ are random Rayleigh distributed variables, i.e.:
\begin{equation}
P(x)  = 2 x e^{-x^2}\,{\rm d}x\, ,
\end{equation}
where $x=\alpha/\alpha_0$ or $v/v_0$ and $\alpha_0,v_0$ are the root-mean-square values. The distribution for $v_i$ follows from the 3-dimensional Maxwell-Boltzmann velocity distribution assumed in the standard halo model (SHM) of DM, while the distribution for $\alpha_i$ follows from considering a random walk under phase averaging~\cite{Foster:2017hbq,Centers:2019dyn}. The normalisation by $\sqrt{2\rho_{\rm loc}}/m$ in Eq.~\eqref{eqn:coherent} ensures the correct average local DM density. The maximum number of draws from the distributions, $i_{\rm max}$, is a free parameter that should be large enough to well sample the distributions. The model Eq.~\eqref{eqn:coherent} models the field as a sum of \emph{coherent} states of definite phase and energy, and corresponds to the coherent state $|\phi\rangle$ under which the expectation value of the field operator obeys the classical equations of motion. Note that $|\phi\rangle$ is not an eigenstate of the number operator, $\hat{N}=\hat{a}^\dagger \hat{a}$.

The energy density of the field is:
\be
\rho_\phi=\frac{1}{2}\dot{\phi}^2+\frac{1}{2}m^2\phi^2\, ,
\ee
which can be interpreted in the particle picture as a number density, $n_\phi= \rho_\phi/m$.

The coherent stochastic model, Eq.~\eqref{eqn:coherent}, and its (statistical) agreement with MFT simulation~\cite{Veltmaat:2018dfz}, is illustrated in Fig.~\ref{fig2}. The MFT simulation solves the classical equation of  motion of the scalar field in the WKB limit:
\be
\phi = \psi e^{imt}+\psi*e^{-imt}\, .
\ee
Clearly visible in the figure is the coherence time, $\tau_c$, over which the number density undergoes $\mathcal{O}(1)$ fluctuations. Note that the coherent stochastic model cannot predict the exact time series of the MFT simulation, only the coherence time and amplitude of the number density fluctuations. On the other hand, since the MFT simulations are performed under the WKB approximation, they do not include the rapid oscillations on time scale $1/m$ shown in the zoom in of the stochastic model in Fig.~\ref{fig2}(b).

A coherent state virialised model like the one expressed in Eq.~\eqref{eqn:coherent} is explicitly assumed in many UBDM direct detection data analyses, including Refs.~\cite{Centers:2019dyn,Masia-Roig:2022net}, and the model is implemented in the \textsc{AxiScan} data analysis software~\cite{Foster:2017hbq}.~\footnote{\url{https://github.com/bsafdi/AxiScan}}

\section{Incoherent particle model} 

Ref.~\cite{Eberhardt:2021iiq} approximate quantum evolution by a sum of coherent MFT simulations with classically distinct phases. It is found during gravitational collapse and virialisation the phases become incoherent in different DM ``streams''. I approximate this incoherent model by taking a sum of independent copies of the coherent stochastic model above. First, we write the WKB enevelope function, $\psi$, in the stochastic model:
\be
\psi = \frac{\sqrt{2\rho_{\rm loc}}}{m} \sum_{i=0}^{i_{\rm max}} \alpha_i \exp \left[ i\left(m\frac{v_i^2}{2}t+\delta_i\right)\right]\, .
\ee
Labelling each stream by the index $k$ define: 
\be
\psi_k = \frac{\sqrt{2\rho_{\rm loc}}}{m} \sum_{i=0}^{i_{\rm max}} \alpha_i \exp \left[ i\left(m\frac{v_i^2}{2}t+\delta_i+\epsilon_{ik}\right)\right]
\, ,
\ee
where $\alpha_i$, $v_i$ and $\delta_i$ are the same random draws as for the classical stochastic model, but we add a new phase $\epsilon_{ik}$ to every term in the sum. The local number density is given by the ensemble average over streams:
\be
n = \frac{1}{2}m\langle|\psi_k|^2\rangle_k\, .
\ee
The number density is shown Fig.~\ref{fig2}(d). Since the phases are completely uncorrelated for a given energy, the stream ensemble average leads to a number density with no coherent time variation, consistent with the results of Ref.~\cite{Eberhardt:2022exf}. Averaging the number density oscillations requires observing many particles from different streams. In the example in Fig.~\ref{fig2}(d), I found that of order 100 draws are required to visibly reduce the oscillations: the precise number necessary will in reality depend on the experimental precision required.

The incoherent and coherent models of DM give consistent predictions when appropriately averaged (see e.g. Ref.~\cite{Hertzberg:2016tal}). In the case of a DM halo simulated using classical fields, the ensemble average in the particle case is recovered from either the angle average of the classical field (as expressed by the radial DM density profile $\rho(r)$, which has no significant time dependence over the coherence time~\cite{Veltmaat:2018dfz}), or by the coherence time average.

\section{Measuring the quantum state of DM} 

\subsection{Preliminaries}

In an experiment, the DM is converted into some observable signal, with rate $\Gamma_{\phi}$, leading to a power:
\be
\mathcal{P} = V \Gamma_{\phi}E_\phi n_\phi\, ,
\label{eqn:power}
\ee 
where $V$ is the effective volume and $E_\phi$ is the particle energy (see below). The specific details of the conversion from DM to signal are not important here, but for example this can occur by inducing electric fields~\cite{1983PhRvL..51.1415S}, magnetization~\cite{2014PhRvX...4b1030B}, effective currents~\cite{2016PhRvL.117n1801K}, or atomic energy level shifts~\cite{Arvanitaki:2014faa}, for scalar, axion, or dark photon DM, depending on details of the DM particle physics model and the specific experiment (see e.g. Refs.~\cite{Adams:2022pbo,Antypas:2022asj}). A reference scale for the power in the case of the QCD axion in a resonant cavity haloscope at around 1 GHz is $10^{-22}\text{ W}$.

The power, $\mathcal{P}$, has a characteristic linewidth, $\Delta \omega/\omega$. In the coherent model, the linewidth can be derived from the time series power spectrum (i.e. Fourier transform) of the coherent classical field. In the incoherent case the linewidth arises from the energy spread, via $E_i=\hbar\omega_i\approx m c^2+\frac{1}{2}mv_i^2$ (restoring $\hbar$ and $c$ units for clarity), with $v_i$ the velocity of an individual particle. In both cases, the linewidth is determined by the Maxwell-Boltzmann distribution of velocities in the SHM, and is of order $\Delta\omega/\omega\approx (v/c)^2\approx10^{-6}$. Linewidth models used, for example, in an axion haloscope analysis such as described in Ref.~\cite{OHare:2017yze,Brubaker:2017rna} are therefore not sensitive to the model of the quantum state of the field presented here. Haloscopes such as ADMX~\cite{ADMX:2021nhd} operating at sensitivity to the QCD axion with $m_a\approx 3\,\mu\text{eV}$ typically scan a single frequency for $T=100\text{ s}$. Operating only in the frequency domain and averaging over many coherence times, the resulting power is insensitive to the DM quantum state.

In many cases the experimental signal of ultralight DM is derived assuming a coherently oscillating classical source~\cite{1983PhRvL..51.1415S,2013JCAP...04..016H,2016PhRvL.117n1801K,Millar:2016cjp,Abel:2017rtm,Schutte-Engel:2021bqm}, however the presence of the signal does not necessarily depend on the coherence of the field. Indeed, in the case of axion DM, it has been shown in detail that the rate $\Gamma_\phi$ agrees between calculations using classical MFT (i.e. axion electrodynamics) and tree level quantum field theory~\cite{Ioannisian:2017srr}. 

\subsection{The Crux of the Argument}

If the DM is in a classical coherent state, then the models outlined above and shown in Fig.~\ref{fig2} predict that the measured power undergoes $\mathcal{O}(1)$ oscillation over the time period $\tau_c$. In the incoherent quantum model, the possibilities are more complex since the state is a superposition of classical states. Measurement projects the quantum state onto the pointer state (in the case of the cat: alive or dead). Measurement can occur in the DM detector, but also occurs due to entanglement with the environment, i.e. decoherence, which occurs for a cat state. The pointer states are those states that remain as pure as possible under decoherence. Ref.~\cite{Allali:2020ttz} showed that DM can be decohered due to gravity in the environment on a short timescale. 

The coherent and incoherent models differ in their prediction for the time series of the measurement power \emph{if the pointer state is not the classical coherent state} (the relevance of pointer states for UBDM was first discussed in Refs.~\cite{Eberhardt:2021okc,Eberhardt:2022rcp,Eberhardt:2022exf}). In this case, the incoherent state can be measured as such, with no time variation of the measurement power over the coherence time~\cite{Eberhardt:2022exf}. Therefore experiments operating at low particle mass that can resolve the time dependence of the signal on the order of $\tau_c$ can distinguish between the two models and pointer states. The fact that the quantum versus classical description of the DM field $\phi$ is encoded in the time evolution of the number density was noted already in Ref.~\cite{Ioannisian:2017srr}. 

Electromagnetism has the pointer state as the coherent classical field, as do Bose-Einstein condensates with certain types of interactions, but not others: coherent states can be considered the natural pointer states for harmonic oscillators coupled linearly to an environment~\cite{PhysRevD.53.7327}. On the other hand, for interactions such as those of UBDM, both in experiments and gravitationally, it is not a priori obvious what the pointer state should be. Measurement can determine what the DM quantum state is by time series analysis over the classical coherence time.

\subsection{Example}

It is helpful to give a concrete example of the proposed measurement, for which we use the CASPEr experiment~\cite{2014PhRvX...4b1030B,2013PhRvD..88c5023G,JacksonKimball:2017elr}, and specifically the analysis model described in Ref.~\cite{Centers:2019dyn}. CASPEr-Gradient proposes to measure the spin-precession of hyperpolarised liquid Xenon nuclei induced by axion-like DM in an external magnetic field, $B_0$. The axion field, $\phi$, acts like an effective magnetic field, $B_{\rm eff}=g \nabla \phi/ \gamma$, where $g$ is the axion-nucleon coupling constant, and $\gamma$ is the gyromagnetic ratio (a particle description exists just as for nuclear magnetic resonance). The effective magnetic field would induce anomalous magnetization in the Xe sample, which is resonantly enhanced when the axion frequency, $\omega$, is equal to the Larmour frequency due to the external field, $\mu_0 B_0$. The experiment aims to measure the induced magnetization using a magnetometer. In the absence of a detection, an upper limit can be set on $g$ at a given value of the resonant frequency, which is scanned by varying $B_0$. 

The relevant timescales are: field oscillation time, $1/m$, coherence time, $\tau_c$, and integration time, $T$. I consider a reference integration time of 100 s. If a detection is made, then an experimental campaign can make $N_{\rm tot}$ measurements in some reasonable time scale, for example 1 year, in order to determine the quantum state of the field. In order for the quantum state of the field to be measurable, we require $T<\tau_c<N_{\rm tot} T$. 

We begin with the case of the coherent state. In order for the MFT model to be valid the integration time should satisfy $T> 1/m$ to average over many oscillations of field, which for our reference parameters corresponds to $m\gtrsim 10^{-17}\text{ eV}$. With $T< \tau_c$ then the stochastic nature of the field over the coherence time becomes important. In particular, Ref.~\cite{Centers:2019dyn} argued that a Bayesian model should marginalise over the unknown phase, leading to a degradation in the limits that can be set on the unknown coupling constant $g$. A deterministic prediction in the coherent model is only made when averaging over a time scale longer than $\tau_c$. The coherent state leads to fluctuations in the signal power for repeated measurements in a campaign if the campaign time $N_{\rm tot} T\gtrsim \tau_c$. 

In the incoherent model of the axion field $T$ should be long enough to measure many different axion events. If the events sample many streams with different phases, then the prediction for $n$ has very small fluctuations, as shown in Fig.~\ref{fig2}. In this model, all $N_{\rm tot}$ repeated measurements should measure the same signal strength. In the event of a detection in the incoherent quantum model, the induced magnetization at frequency $\omega$ will have no time dependence, and in particular no time variation between successive measurements even if $N_{\rm tot}T\gtrsim \tau_c$. 

Since one does not know, a priori, whether the DM is in a coherent or incoherent state, or what the UBDM pointer state is, it is therefore necessary in the event of any detection to make multiple measurements for an extended time period $N_{\rm tot}T>\tau_c$  in order to break the degeneracy between the unknown phase in the coherent state model, and the inferred value of $g$. Measurement in the time domain is sensitive to the phase coherence of the UBDM.

\section{Conclusions} The possible quantum state of UBDM has provoked debate and discussion for decades. Any departure from standard MFT may change the astrophysical and cosmological phenomenology of UBDM, in particular that based on fluctuations on the coherence scale~\cite{Hui:2016ltb,Marsh:2018zyw,2019ApJ...871...28B,El-Zant:2019ios}, and affects the inference of constraints derived from experiment~\cite{Centers:2019dyn}. Cosmological simulations of UBDM beyond MFT are extremely challenging, and have only recently started to be developed~\cite{Lentz:2019xcr,Eberhardt:2021iiq}. Even so, current understanding of UBDM does not specify the pointer state. Thus, I have considered whether and how the quantum state of UBDM can be measured in the laboratory.

If an experiment can be carried multiple times over the coherence time then in a coherent state model the observed signal should fluctuate, while it will not in an incoherent model if the pointer state is not the classical coherent state. Such a measurement is feasible for $\tau_c$ less than around one year, or $m\gtrsim 10^{-17}\text{ eV}$. To observe the time variation the signal should be strong enough to measure on integration times shorter than $\tau_c$, which for a given experiment limits the range of possible UBDM coupling constants where coherence can be measured. Taking a benchmark value of 100 seconds of measurement time suggests an upper limit of where this measurement can be performed of around $m\approx 10^{-11}\text{ eV}$, although this is by no means definitive and further exploration on a case-by-case basis of experiments is necessary. This simple example demonstrates how the issue of the DM field quantum sate can be resolved experimentally if ultralight DM is detected in the laboratory. I hope this stimulates further exploration of the possible phenomenological consequences of the DM quantum state.

\emph{Acknowledgements} I acknowledge useful discussions with Tom Abel, Andrew Eberhardt, Mark Hertzberg, and Michael Kopp, and with participants of the ``Wavy Dark Matter Summer'', 2022. I thank the organisers, especially Arne Wickenbrock and Dima Budker, for a highly engaging program and warm hospitality. I am supported by an Ernest Rutherford Fellowship from the Science and Technologies Facilities Council (UK). I made use of the open source software \textsc{matplotlib}~\cite{matplotlib}, \textsc{numpy}~\cite{numpy}, and \textsc{scipy}~\cite{scipy}. I am indebted to Jan Veltmaat for supplying the simulation results of Ref.~\cite{Veltmaat:2018dfz}.

\bibliographystyle{h-physrev3.bst}

\bibliography{axion}

\end{document}